\documentclass[aps,prd,twocolumns,preprintnumbers,showpacs,amsmath,amssymb]
{revtex4}

\usepackage{graphicx}
\usepackage{dcolumn}
\usepackage{bm}

\begin{document}

\title{A few brief comments to arXiv:1005.3315 and arXiv:1005.3321 }

\author{V. Gogokhia}
\email[]{gogohia@rmki.kfki.hu}

\affiliation{HAS, CRIP, RMKI, Depart. Theor. Phys., Budapest 114,
P.O.B. 49, H-1525, Hungary}

\date{\today}
\begin{abstract}

\end{abstract}

\pacs{ 11.15.Tk, 12.38.Lg}

\keywords{}

\maketitle



In the above-mentioned papers \cite{1,2} the authors use the so-called ladder-rainbow truncation scheme
for the solution of the Dyson-Schwinger equation for the quark propagator. The same approximation is used in the
Bethe-Salpeter bound state equation in order to investigate soft and hard scale QCD dynamics in mesons.
In this connection we would like to make a few general remarks, even not discussing that their decomposition
of the effective running coupling is neither exact nor unique. 
 
\vspace{3mm}

In our papers \cite{3,4} it has been explicitly shown in the most general form that this truncation scheme is inconsistent with the color gauge invariance/symmetry of QCD. The reason of the inconsistency 
is that the color charges interaction is always present in QCD, and this approximation
is not able to correctly take this interaction into account. Such
kind of the interaction is absent in QED, and therefore this approximation works
there (in more detail the comparison between ladder-rainbow approximations in QCD and QED is
discussed in \cite{4}). 

\vspace{3mm}

The ladder-rainbow approximation is definitely wrong for light quarks and becomes
trivial for heavy quarks, when the response of the QCD vacuum can be neglected, indeed.
So the general conclusion is that all the results based on the ladder-rainbow truncation scheme to QCD should be abandoned or, at least, re-considered.

\vspace{3mm}

In \cite{5} it has been made an attempt to claim that our proof of the inconsistency of the
ladder-rainbow approximation to QCD was incorrect. However, in \cite{6} we have shown that their claim was
based on a derivation which contained a mathematical mistake, namely the unjustified
change of variables in the divergent (though regularized) integrals.

\end{document}